\documentclass{article}
\usepackage{arxiv}

\usepackage[T1]{fontenc}
\usepackage[utf8]{inputenc}
\usepackage{times}

\usepackage{graphicx}
\graphicspath{{figures/}}
\DeclareGraphicsExtensions{.pdf,.png,.jpg} 

\usepackage{amsmath,amsfonts}
\usepackage{booktabs}
\usepackage{array}
\usepackage{tabularx}
\usepackage{makecell}
\usepackage[caption=false,font=normalsize,labelfont=sf,textfont=sf]{subfig}
\usepackage{stfloats}
\usepackage{algorithm}
\usepackage{algorithmic}

\usepackage{nicefrac}
\usepackage{microtype}
\usepackage{textcomp}
\usepackage{verbatim}
\usepackage{url}
\usepackage{doi}

\usepackage[numbers,sort&compress]{natbib}

\title{A Survey on Agentic Service Ecosystems: Measurement, Analysis, and Optimization}

\newif\ifuniqueAffiliation
\uniqueAffiliationtrue

\author{
    Xuwen Zhang \\
    College of Intelligence and Computing \\
    Tianjin University \\
    \texttt{1023244032@tju.edu.cn} \\
    \And
    Xiao Xue\thanks{These authors contributed equally to this work and are corresponding authors. Correspondence to: \texttt{jzxuexiao@tju.edu.cn}, \texttt{shelicy@hainanu.edu.cn}.} \\
    College of Intelligence and Computing \\
    Tianjin University \\
    \texttt{jzxuexiao@tju.edu.cn} \\
    \And
    Xia Xie\footnotemark[1] \\
    School of Computer Science and Technology \\
    Hainan University \\
    \texttt{shelicy@hainanu.edu.cn} \\
    \And
    Qun Ma \\
    College of Intelligence and Computing \\
    Tianjin University \\
    \texttt{1023244018@tju.edu.cn} \\
    \And
    Xiangning Yu \\
    College of Intelligence and Computing \\
    Tianjin University \\
    \texttt{yxn9191@gmail.com} \\
    \And
    Deyu Zhou \\
    School of Software, Shandong University \\
    \texttt{zhoudeyu@mail.sdu.edu.cn} \\
    \And
    Yifan Wang \\
    College of Intelligence and Computing \\
    Tianjin University \\
    \texttt{yiifanwaang@163.com} \\
    \And
    Ming Zhang \\
    Faculty of Environment, Science and Economy \\
    University of Exeter \\
    \texttt{mz427@exeter.ac.uk} \\
}

\hypersetup{
  pdftitle={A SURVEY ON Agentic SERVICE ECOSYSTEMS},
  pdfsubject={cs.AI},
  pdfkeywords={Service ecosystem, emergence of swarm intelligence, measurement, analysis, optimization},
}

\begin{document}
\maketitle

\begin{abstract}
The Agentic Service Ecosystem consists of heterogeneous autonomous agents (e.g., intelligent machines, humans, and human–machine hybrid systems) that interact through resource exchange and service co-creation. These agents, each with distinct behaviors and motivations, exhibit autonomous perception, reasoning, and action capabilities, which increase the system’s complexity and render traditional linear analysis methods inadequate. Swarm intelligence—with its characteristics of decentralization, self-organization, emergence, and dynamic adaptability—offers a novel theoretical lens and methodology for understanding and optimizing such ecosystems. However, current research, due to fragmented perspectives and differences among ecosystems, fails to comprehensively capture the complexity of swarm intelligence emergence in Agentic contexts. The lack of a unified methodology further limits the depth and systematic nature of the research. This paper proposes a swarm intelligence emergence analysis framework tailored to Agentic Service Ecosystems, encompassing three steps—measurement, analysis, and optimization—to reveal the cyclical mechanisms and quantitative criteria that foster swarm intelligence emergence. By reviewing existing technologies, the paper analyzes their strengths and limitations, identifies unresolved challenges, and demonstrates how this framework not only offers theoretical support for the formation of swarm intelligence in agentic ecosystems but also provides actionable methods for real-world applications.
\end{abstract}


\section{Introduction}
With 
the rapid advancement of new-generation information technologies such as service computing, cloud computing, the Internet of Things, and blockchain, enterprises and organizations are undergoing a service-oriented transformation[1], encapsulating various business elements (including applications, data, and resources) into invokable services [2]. Service systems are gradually evolving into Agentic Service Ecosystems, distributed environments composed of heterogeneous autonomous service agents (humans, intelligent robots, and human–machine integrated systems) that can perceive, reason, and act in pursuit of individual and collective goals[3]. These service agents, with their increasing levels of intelligence, exhibit characteristics of collective intelligence within the ecosystem [4]. The interactions among agents with varying levels of intelligence form a complex interdependent network, demonstrating self-organizing properties[5]. Due to these two characteristics, early service ecosystems were functionally simplistic, primarily pursuing efficiency and stability, and only required consideration of linear input-output relationships. With the development of intelligence and diversification, service ecosystems have gradually evolved into distributed, dynamic, and intelligent complex systems, exhibiting nonlinear characteristics such as feedback loops and emergent behaviors. Swarm intelligence, through simple rules and local interactions among individuals, gives rise to complex global behaviors [6]. Its collective intelligence, self-organization, and nonlinearity pose new challenges for the governance of service ecosystems.

The governance of Agentic Service Ecosystems is increasingly regarded as a highly challenging task, primarily due to the collective intelligence aspect. As the intelligence of agents increases, the phenomenon of swarm intelligence becomes further complicated. Individual agents may prioritize their own interests, leading to conflicts with collective goals. For instance, human agents may tend to maximize personal benefits, while machine agents may overly focus on efficiency optimization, neglecting the overall impact on the system. Secondly, the self-organizing nature among agents with different levels of intelligence stems from the intricate interdependencies among stakeholders, manifesting as a complex network. Traditional linear governance models are ill-equipped to handle such dynamic, nonlinear, and uncertain interactions. In summary, the governance of service ecosystems faces the following three major challenges:

\textbf{Evolutionary Complexity (Difficulty in Measurement):}The evolution of service ecosystems involves the development of the overall functionality of multi-agent behaviors and their associated structures. The lack of an appropriate system for measuring service effectiveness (such as comprehensive capability assessment) makes continuous improvement and governance challenging [7].

\textbf{Relational Complexity (Difficulty in Analysis): }The evolution of service ecosystems, from multi-agent behaviors to the evolution of overall functionality, challenges the effectiveness of traditional hierarchical management. Network interconnectivity and positive feedback amplification mean that governance strategies may trigger a cascade of "1+1<2" or "1+1>2" chain reactions. Existing analytical methods predominantly focus on "description" rather than "explanation" [8].

\textbf{Regulatory Complexity (Difficulty in Optimization): }The misalignment between individual interests and the overall system benefits may lead to conflicts in order and stability, challenging traditional governance models. Governance strategies need to balance the pursuit of individual interests with system development to ensure the vitality of the system [9].
\begin{figure}[!t]
\centering
\includegraphics[width=3.5in]{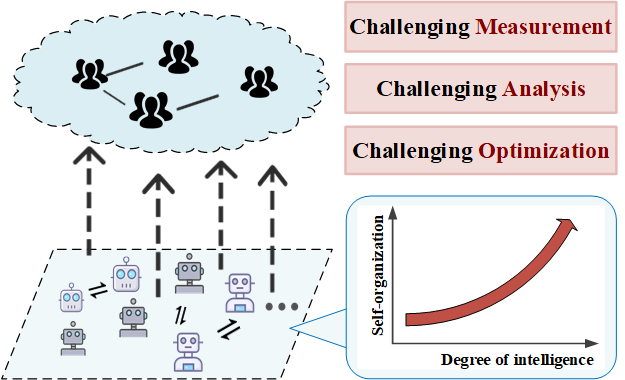}
\caption{The overall structure of the main content of this paper.}
\label{fig_1}
\end{figure}

In previous research on service ecosystems, the mechanisms of swarm intelligence formation and its role in promoting collaborative processes and value co-creation have been demonstrated to hold significant scientific importance. Existing studies on swarm intelligence in service ecosystems have largely focused on specific areas such as the impact of service delivery, the enhancement of customer experience, and technological integration. During the synthesis of individual intelligent agents into a collective, they enable adaptive responses within the service environment, thereby fostering the formation of a collaborative ecosystem [10]. The dynamic interactions among various service stakeholders, characterized by mutual interactions and dependencies among intelligent services, construct a network-driven service ecosystem that facilitates value co-creation [11]. Although research pathways in swarm intelligence have validated its importance in enhancing collective decision-making and collaboration, the transformation of individual contributions into measurable collective outcomes remains a significant challenge.

Existing studies have yet to adequately address the difficulty of quantifying the transition from individual to collective effects in the process of swarm intelligence formation within service ecosystems[12]. The primary reason for this is the lack of a methodological framework for understanding the mechanisms of swarm intelligence formation. To this end, this paper proposes a swarm intelligence emergence analysis framework encompassing three stages: "measurement, analysis, and optimization," aiming to reveal the cyclical mechanisms and measurement criteria that promote swarm intelligence emergence in service ecosystems. For each step, this paper discusses its objectives and challenges and points out future research directions. The core contribution of this paper lies in providing theoretical support and practical guidance for swarm intelligence formation in service ecosystems through this methodological framework.

The remaining structure of this paper is organized as follows: Chapter 2 introduces the three stages of service ecosystems and their correspondence with swarm intelligence; Chapter 3 constructs a theoretical model of swarm intelligence formation mechanisms; Chapter 4 explores measurement methods for service ecosystems; Chapter 5 investigates analytical methods for service ecosystems; Chapter 6 analyzes optimization methods for decision-making in service ecosystems; and Chapter 7 concludes the paper and outlines future research directions.

\section{The Three Stages of Swarm Intelligence Development in Service Ecosystems}
With the continuous advancement in the intelligence level of service agents, to better govern service ecosystems, it is necessary to categorize swarm intelligence based on the distinctions in heuristic intelligence according to their developmental stages. The intelligent development of service ecosystems can be divided into three phases:

The first phase is grounded in biologically inspired intelligence, focusing on integrating theoretical methods with specific domains to address practical problems, thereby enhancing the operational efficiency and adaptability of service ecosystems.

The second phase is based on socially inspired intelligence, leveraging mobile internet and human network systems to optimize resource allocation and service processes. It aims to provide personalized services and decision support through intelligent and efficient means to meet user demands.

The third phase is rooted in brain-inspired intelligence, emphasizing collaboration and integration between individuals and groups. The core objective is to resolve conflicts between individual benefit maximization and collective benefit maximization.

Through the evolution of these three phases, service ecosystems have progressively achieved a leapfrog development from theory to practice, from local optimization to global collaboration, and from basic services to intelligent decision-making.
\subsection{First Stage of the Service Ecosystem: Bio-inspired Swarm Intelligence.}
In 1989, Beni et al. first introduced the concept of "Swarm Intelligence" in their article "Swarm Intelligence in Cellular Robotic Systems"[13]. This concept was inspired by natural phenomena observed in biological systems, such as the coordinated foraging behavior of ant colonies and the flocking dynamics of bird groups. Subsequently, the intelligence exhibited at the level of lower social biological groups has commonly been referred to as swarm intelligence. In early service ecosystems, the intelligence levels of service agents were relatively similar, and the intelligence of individual service agents often consisted of homogeneous entities, exhibiting similar programming structures and decision-making processes[14]. This consistency among agents simplified communication and coordination within the service ecosystem, facilitating direct collective behavior [15].

This biologically inspired swarm intelligence phenomenon manifests in service ecosystems as a method to enhance operational efficiency through local interactions among agents, akin to the foraging behavior of ant colonies, which operate effectively without centralized control. Swarm Intelligence (SI) has emerged as a crucial tool for optimizing service systems, particularly demonstrating significant advantages in selecting and organizing services from vast service pools to meet specific user requirements [16]. In service composition, SI algorithms can effectively address environmental constraints and objectives. For instance, the hybrid PSO algorithm proposed by Dahan et al., which incorporates genetic operators, enhances the ability to meet Quality of Service (QoS) requirements, showcasing the synergistic potential of SI with traditional optimization strategies [17]. Furthermore, SI has made strides in enhancing communication and collaboration among agents, as evidenced by research on artificial fish swarm algorithms and dolphin swarm algorithms, which significantly improve the robustness and efficiency of service discovery and composition [18].
\subsection{Second Stage of the Service Ecosystem: Social-Inspired Collective Intelligence.}
In 2017, Chai Yueting's proposal of "Crowd Science and Engineering: Concept and Research Framework"(CSE) garnered widespread attention in academia [19]. This framework leverages technological interconnectivity and human collaboration to harness collective intelligence, offering innovative pathways for social operations and management. The second phase of service ecosystems, characterized by socially inspired intelligence, emphasizes the role of social interaction and collaboration in decision-making. By integrating human cognition with computational capabilities, it provides novel approaches to addressing complex societal challenges. At the core of socially inspired intelligence is the tripartite integration of humans, machines, and objects, enabling efficient scheduling and collaborative operations among intelligent agents. This phase focuses on interdisciplinary developments with technologies and concepts such as the Internet of Things (IoT), the internet, big data, collaborative computing, and human-machine-object integration, thereby generating synergistic effects where "the whole is greater than the sum of its parts."

During the socially inspired intelligence phase, the convergence of large-scale data processing, human-computer interaction, and intelligent applications is experiencing significant growth [20]. The digital economy, as a cornerstone of modern business dynamics, extensively utilizes data to enhance public services, including health and education, thereby achieving better governance and more effective interactions between governments and citizens [21]. Through the adoption of collaborative computing and mobile crowdsourcing frameworks, fields such as intelligent transportation and healthcare are undergoing revolutionary transformations in service delivery and operational efficiency [22]. In this phase, service ecosystems are capable of providing more precise personalized service recommendations and intelligent decision support, marking a significant breakthrough in service intelligence.
\subsection{Third Stage of the Service Ecosystem: Brain-Inspired Crowd Intelligence.}
Li Wei and colleagues, in China's new-generation artificial intelligence strategic plan, refer to the intelligence that emerges from the stimulation of individual wisdom and the aggregation of collective wisdom in large-scale complex groups as “Crowd Intelligence” [23]. This concept integrates the previously proposed “Swarm Intelligence” and “Collective Intelligence”, aiming to establish a unified research framework that accommodates both low-intelligence and high-intelligence individuals, thereby advancing crowd intelligence as a new paradigm for solving large-scale complex problems in open and uncertain environments. Crowd Intelligence emphasizes collaboration and integration between individuals and groups, providing more comprehensive and efficient solutions to complex problems. Since the release of ChatGPT in 2022, the general language understanding, generation, and emergent capabilities of Large Language Models (LLMs) have enabled users to interact with systems. The third phase of service ecosystems has since evolved to focus on intelligent systems with near-human language understanding and cognitive abilities. In large-scale data processing and complex decision-making, this phase leverages brain-inspired parallel computing, associative memory, and other characteristics to uncover the latent wisdom of groups, thereby enabling effective solutions to complex problems [24].

In this brain-inspired intelligence phase, service ecosystems are trained on vast datasets to acquire "human-like" response capabilities. As the intelligence of agents increases, these capabilities influence individual behaviors through social learning mechanisms, allowing individuals to benefit from group dynamics and optimize decision-making. This promotes alignment between individual behaviors and group dynamics, driving collective improvement. The core challenge of this phase lies in the conflict between individual and collective benefit maximization, reflecting the deep-seated contradictions between social dynamics and collective behavior.
\begin{figure}[ht]
    \centering
    \includegraphics[width=\linewidth]{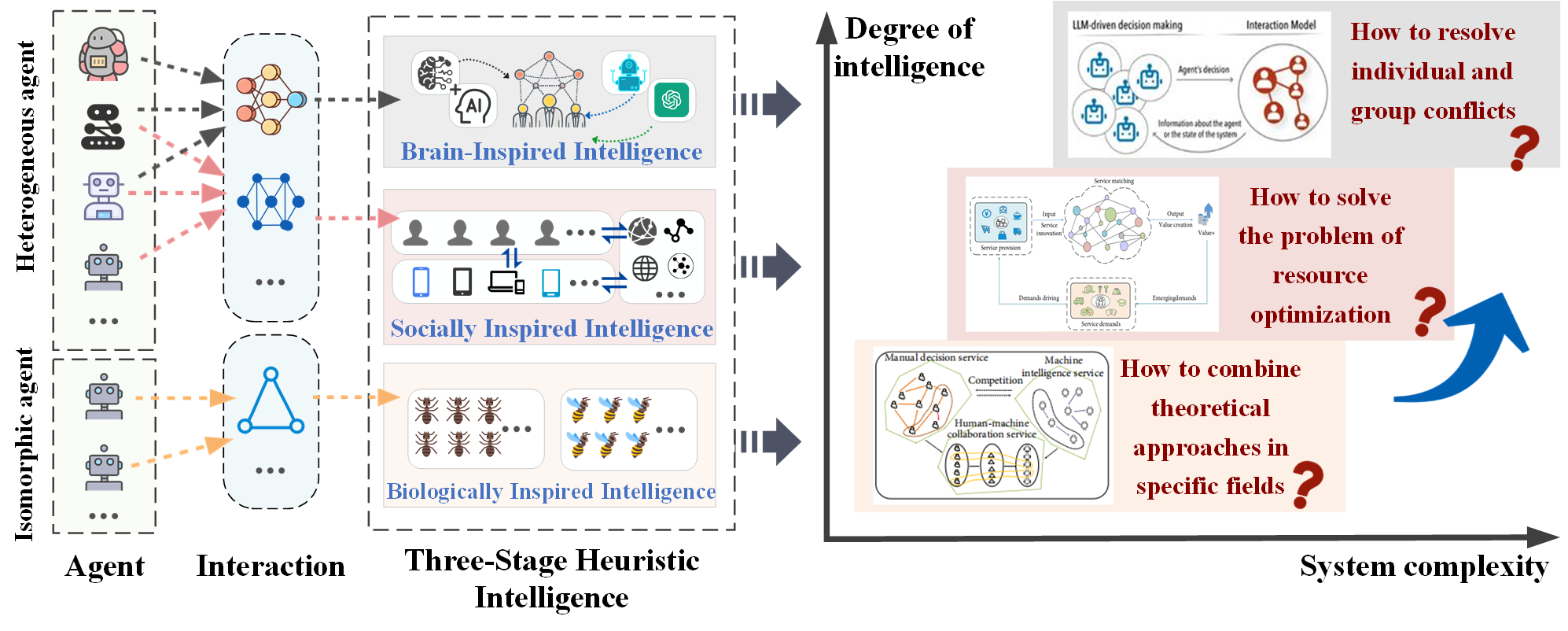} % 或 figures/2.png
    \caption{Evolution of swarm intelligence at three stages in a service ecosystem.}
    \label{fig:evolution_stages}
\end{figure}

\subsection{Evolution and Quantification Challenges of Swarm Intelligence in Service Ecosystems}
The concept of swarm intelligence originates from observations of natural systems. Individual motivation is the first step in the evolution of swarm intelligence, where individuals within a group independently explore problems, acquire information, and contribute to complex tasks through parallel exploration, task decomposition, and solution finding. This process involves changes in individual behaviors or states, necessitating modeling studies to understand the dynamics of each agent during this phase. The typical structure of an agent includes perception, decision-making, action, and optimization. Agent models can be customized to address specific problems. In the evolutionary process of swarm intelligence, individual motivation serves as the initial step[25].

The formation of collective intelligence occurs as the group expands, requiring the integration of individual achievements to accomplish overarching task objectives. This necessitates effective model integration at the team level to coordinate collective behaviors. A defining characteristic of swarm intelligence is the intelligent interaction and communication among individuals, enabling the group to self-organize and transition from diverse states to stable and orderly states.

The aggregation of individuals into groups and the formation of collective intelligence is a multifaceted process that encompasses various dimensions of society, including social learning, communication technologies, and psychological factors. Among these, social learning is one of the key mechanisms. According to social learning theory, individuals acquire knowledge and behaviors from others through observation and interaction. The formation of collective intelligence relies on social interactions among members, enabling them to draw from and expand upon each other's ideas. Belova E et al. emphasize the significant role of social interaction in knowledge accumulation [26], while Bazazi et al. further note that groups can achieve collective intelligence that surpasses individual capabilities by integrating independently acquired information [27]. Secondly, communication technologies play a pivotal role in group integration, and their importance cannot be overstated. The advent of digital communication tools has not only revolutionized the dynamics of team interactions but also facilitated efficient collaboration across geographical and cultural boundaries, thereby significantly enhancing overall performance and innovation capabilities. The psychological aspects of group dynamics also play a crucial role in the process of group cohesion. When individuals become part of a group, their perceptions and behaviors undergo changes. Momennejad et al. illustrate this phenomenon, finding that individuals in like-minded groups tend to converge on shared memories and interpretations of events, influenced by their group identity [28]. This convergence of perspectives can lead to a stronger sense of group identity, further reinforcing cohesion and collective intelligence.

However, at present, the impacts of social learning behaviors, communication technologies, and psychological influences at both individual and group levels are difficult to quantify. In constructing complex service ecosystems, exploring the intrinsic mechanisms of data fusion, interaction, and intelligent decision-making among agents during the generation of swarm intelligence is crucial. This process urgently requires an intrinsic mechanism to provide quantifiable feedback based on data and information. Logically, these intrinsic mechanisms can be summarized as a cyclical model of measurement, analysis, and optimization. Specifically, in the process of generating intelligence from group behaviors, accurate measurement, in-depth analysis, and optimization of individual and group behaviors are necessary to facilitate the effective generation of system intelligence.
\section{A Framework for Research Perspectives on Swarm Intelligence-Oriented Service Ecosystems}
The complexity of service ecosystems arises not only from the continuous enhancement of collective intelligence but also from the following two factors: first, their intrinsic mechanisms and governing principles are rarely limited to one or two; instead, multiple internal mechanisms and laws often operate simultaneously. Second, complex systems are typically open systems, subject to numerous external forces. When will these external forces occur? What impact will they have—positive, negative, or sometimes positive and sometimes negative? These questions are often difficult to determine in advance.

Against this backdrop, to clearly delineate the organizational structure and evolutionary processes of service ecosystems, this paper proposes a constructive model comprising three steps: "measurement," "analysis," and "optimization." The aim is to elucidate the theory of swarm intelligence formation, leveraging this theoretical framework to design more efficient artificial collective intelligence systems. The constructive model for the formation mechanism of swarm intelligence in service ecosystems is illustrated in Figure 3 below:

\begin{figure*}[ht]
    \centering
    \includegraphics[width=0.8\textwidth]{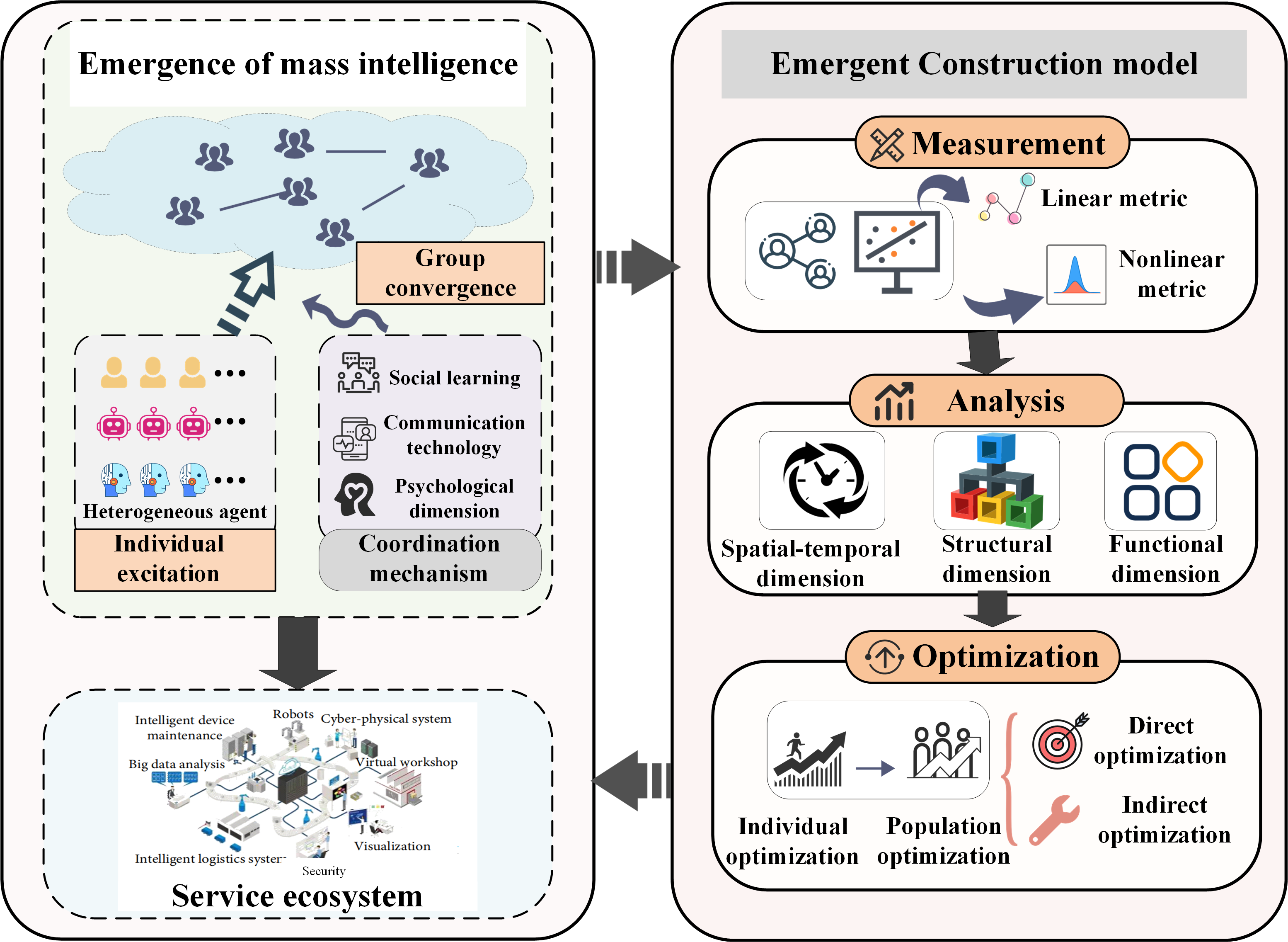} 
    \caption{A constructive model for guiding the formation mechanism of swarm intelligence in a service ecosystem.}
    \label{fig:constructive_model}
\end{figure*}

\subsection{Difficulty in Measurement}
In service ecosystems, the complexity of measurement is increasingly pronounced, primarily due to the dynamic behaviors of the system and the inherent complexity of swarm intelligence [29]. Traditional linear measurement methods, which focus solely on the direct relationship between inputs and outputs, fail to effectively capture the intricate interdependencies and nonlinear interactions in modern service ecosystems [30]. The emergent properties exhibited by these systems often cannot be predicted by analyzing individual components in isolation [31]. When users seek solutions that meet their needs from a vast pool of services, measurement must be conducted at both individual and collective levels. At the individual level, measurement is typically based on personal needs and experiences, while at the collective level, it is influenced by consensus and shared requirements.

Several issues regarding measurement remain unresolved:
\begin{itemize}
    \item First, innovative services lack mature measurement methods and standards, making it difficult to accurately assess their value and determine their future development potential.
    \item Second, it is challenging to balance the interests of multiple stakeholders and ensure that all parties receive appropriate value returns.
    \item Measurement methods need to consider perspectives at both individual and group levels to capture diverse needs and preferences.
\end{itemize}

Addressing measurement challenges in service ecosystems requires a nuanced understanding of the interplay between individual and collective perspectives, the complexity of ecosystem interactions, and the diverse interests of the various stakeholders involved. Future research should focus on identifying comprehensive measurement approaches capable of accommodating these complexities, balancing the differing interests between individual and collective viewpoints, while ensuring that all parties derive fair value from the services provided.
\subsection{Difficulty in Analysis}
Analyzing service ecosystems has become increasingly challenging, primarily due to the self-organizing nature of swarm intelligence. This characteristic leads to complex higher-order dynamics that are difficult to quantify using traditional methods [32]. Simultaneously, traditional hierarchical management structures struggle to adapt to the dynamic changes and emergent behaviors brought about by swarm intelligence [33]. Furthermore, individual adaptive changes and the secondary emergence of collective behaviors further exacerbate the complexity of analysis [34]. In service ecosystems, the objects of analysis include various types of services, and the overarching goals of analysis are multifaceted: on one hand, quantifying the value of service ecosystems can assist decision-makers in making informed choices; on the other hand, integrating the perspectives and knowledge systems of different participants within the ecosystem can promote its evolution.

The analysis highlights two major challenges currently faced:
\begin{itemize}
    \item First, the challenge in analyzing service ecosystems lies in coordinating all participants to achieve a unified collaborative service objective.
    \item Second, there is currently a lack of comprehensive analysis and understanding of formal verification methods used to validate the correctness of service compositions in communication networks.
\end{itemize}

Addressing the analytical challenges in service ecosystems requires the capability to capture the multifaceted nature of these ecosystems. It is essential to adopt a swarm intelligence perspective and conduct systematic analysis by integrating temporal-spatial, structural, and functional dimensions. Future research should focus on the self-organizing characteristics of swarm intelligence, combining multiple dimensions to uncover the complexity of the system and its underlying principles.
\subsection{Difficulty in Optimization}
In modern service systems, optimization faces numerous challenges, primarily stemming from the need for groups within service ecosystems to adapt to ever-changing environments and ensure appropriate resource allocation and utilization [35]. Firstly, the shift toward decentralization in service systems has distributed decision-making authority across multiple agents, each possessing autonomous optimization capabilities [36]. Secondly, individual agents often prioritize their own interests: human agents tend to favor short-term gains, while machine agents excessively pursue efficiency at the expense of system stability [37]. This conflict between individual and collective interests further hinders system optimization [38]. Additionally, as the intelligence level of agents increases, traditional hierarchical management models struggle to cope with diverse participants and complex network relationships. The lack of effective governance and regulation mechanisms makes optimization even more difficult to achieve.
\begin{itemize}
    \item One of the primary challenges in service ecosystems is the complexity and dynamic nature of these systems.
    \item Meeting heterogeneous and customized optimization needs through reasonable optimization measures.
\end{itemize}

As service ecosystems continue to evolve, there is an urgent need to develop adaptive optimization techniques that align with the dynamic characteristics of these ecosystems. Future research should also focus on tailoring optimization methods to accommodate the unique features of different service domains. This not only requires an understanding of the specific needs of each domain but also the development of flexible optimization frameworks capable of seamlessly adapting to diverse environments.
\section{Measurement in Service Ecosystems}
In the early stages of service ecosystem development, measurement methods primarily relied on linear metrics, which focused on direct causal relationships and single variables, emphasizing simplicity and efficiency. However, as service ecosystems have evolved into complex systems, particularly driven by swarm intelligence, the limitations of linear metrics have become increasingly apparent. Consequently, measurement methods are transitioning from linear to nonlinear approaches. Nonlinear metrics are capable of more accurately capturing the dynamic behaviors and adaptability of systems.

This chapter discusses how advancements in swarm intelligence have spurred the urgent need for nonlinear metrics. This shift enables a more comprehensive understanding of interactions and dynamic behaviors within complex service ecosystems, providing a more precise framework for system evaluation. The evolution of measurement approaches is illustrated in Figure 4 below:
\begin{figure}[!t]
    \centering
    \includegraphics[width=0.75\linewidth]{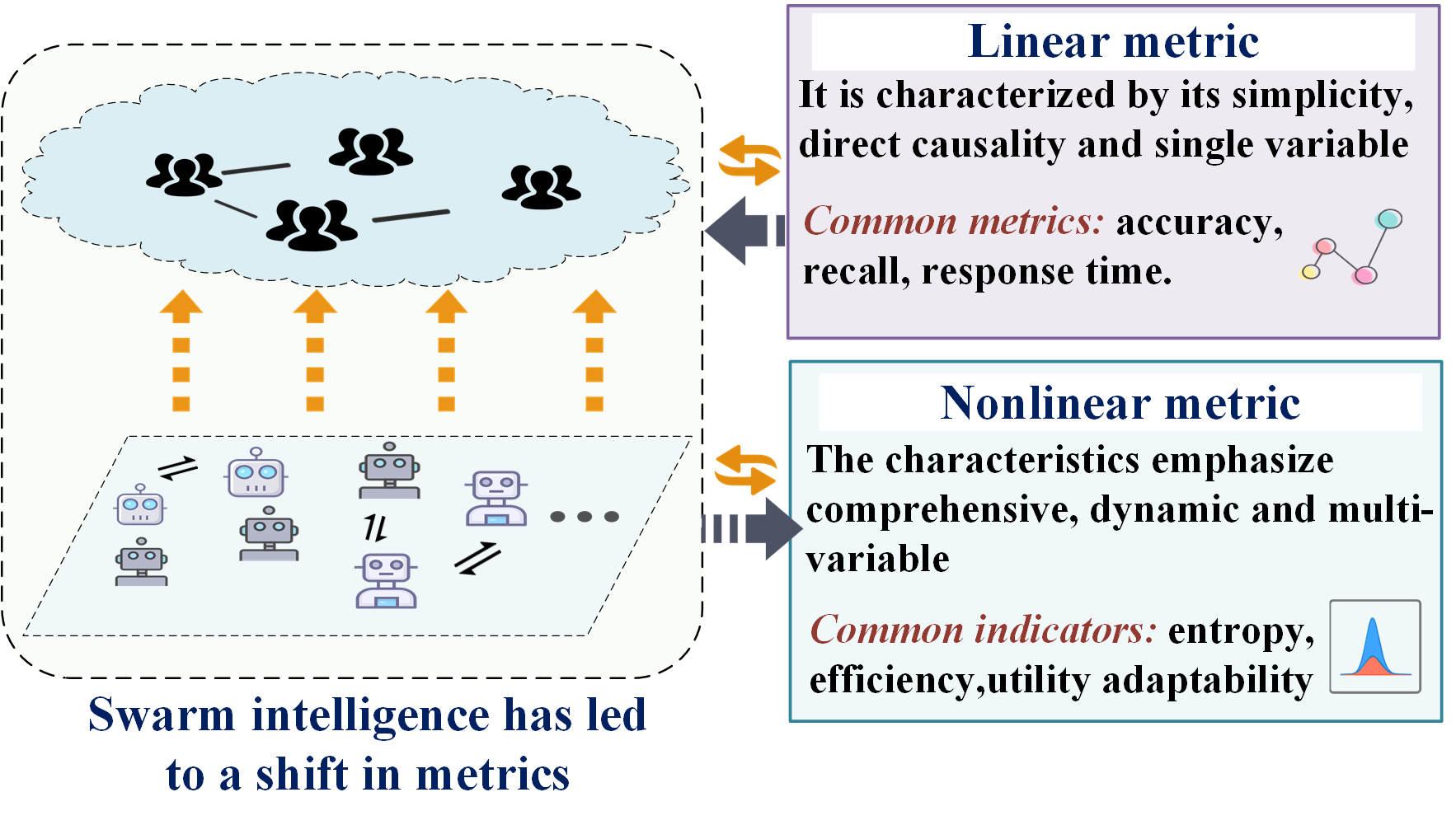} 
    \caption{Linear and nonlinear measurement objective diagram.}
    \label{fig_1}
\end{figure}

\subsection{Linear-Based Measurement}
Traditionally, the evaluation of service ecosystems has relied on linear metrics such as accuracy, recall, and response time, which provide a simplified assessment framework. These metrics illustrate the linear relationship between service inputs and outputs, making them prevalent across various service domains for evaluating system effectiveness and efficiency [39]. However, as service interactions become increasingly complex and dynamic, the limitations of these traditional metrics have become more apparent. They struggle to encapsulate emergent behaviors caused by nonlinear interactions among multiple variables, thereby failing to deliver a comprehensive and accurate assessment of service performance.

Moreover, within the context of service ecosystems, the role of group consensus mechanisms, particularly trust and reputation, is indispensable. Trust not only facilitates smoother service interactions but also reduces transaction costs and fosters collaboration among stakeholders [40]. Similarly, reputation mechanisms offer critical insights into the reliability and quality of service providers, thereby guiding consumer choices [41]. Nevertheless, assessing trust and reputation presents significant challenges, such as information lag and the potential for misleading information, which can distort evaluations and undermine the credibility of service providers [42]. These issues highlight the shortcomings of existing evaluation frameworks, which may inadequately address the complexities of modern service environments characterized by dynamic interactions and emergent properties [43].

To address these challenges, future research should focus on developing robust evaluation frameworks that integrate quantitative metrics with qualitative analysis. Such a hybrid approach would provide a more comprehensive understanding of nonlinear relationships and dynamic changes within service ecosystems. The pursuit of innovative models that combine traditional measures with new evaluation methods could lead to better management practices in the service sector, adapting to the evolving nature of service delivery. Ultimately, while traditional metrics serve as a foundation, they are clearly insufficient for addressing the complexities of contemporary service ecosystems, necessitating more adaptive and comprehensive evaluation frameworks [44].
\subsection{Swarm Intelligence Drives the Shift from Linear to Nonlinear Measurement}
The rapid advancement of swarm intelligence is profoundly reshaping the evaluation landscape of service ecosystems. Traditional linear metrics, such as customer experience (e.g., reliability, responsiveness) and objective performance indicators (e.g., service accessibility, time cost), while historically effective tools for individual-level service assessment, are increasingly inadequate in the face of the complexity and dynamism introduced by swarm intelligence[45]. Swarm intelligence introduces new dimensions where individual behaviors are significantly influenced by group interactions, making traditional linear models ill-suited to capture the uniqueness and diversity of user needs. Consequently, there is an urgent need to develop flexible nonlinear models that better adapt to and reflect the dynamic changes in these complex environments [46-48].

At the individual level, the limitations of traditional metrics are particularly evident. These metrics fail to adequately incorporate the nonlinear interactions arising from collective behavioral elements in swarm intelligence systems. Modern service experiences are no longer solely dependent on direct interactions between service providers and customers but are deeply influenced by community feedback and peer interactions[49]. This shift necessitates more comprehensive and adaptive data collection and integration methods to address the complex dynamics of cross-system coordination [50]. However, this complexity also makes measurement accuracy more challenging [51]. Therefore, the ability to interpret these new patterns is crucial for achieving meaningful evaluation outcomes in modern service ecosystems [52].

At the group level, the decentralized decision-making and self-organizing behaviors of swarm intelligence further amplify nonlinear interactions within service ecosystems. These interactions pose significant challenges to existing static and one-dimensional evaluation frameworks. As stakeholders in service ecosystems encompass diverse groups such as individual users, service providers, and policymakers, developing evaluation frameworks that accommodate multi-level metrics and adapt to the perspectives of different stakeholders becomes particularly important [53]. This complexity underscores the importance of hierarchical data integration and sophisticated modeling to comprehensively understand the system's overall behavior [54-56].

As service ecosystems increasingly rely on swarm intelligence, traditional linear metrics are gradually losing their dominance. Instead, innovative nonlinear frameworks that integrate emerging technologies and insights into collective behaviors will become key to accurately assessing performance and effectiveness in dynamic environments. This shift reflects a broader paradigm change, recognizing the importance of complex adaptive systems in managing cross-sector services and progressively abandoning outdated methods that fail to meet the challenges of modern service environments [57-59].
\subsection{Nonlinear-Based Measurement}
The shift from traditional linear metrics to nonlinear metrics is crucial for accurately capturing the dynamic interactions between individuals and groups within ecosystems. Traditional metrics often struggle to encompass the complexity of these interactions and their emergent effects on the overall system performance. The limitations of single metrics necessitate the adoption of comprehensive indicators, which serve as core tools for more holistically measuring system performance through nonlinear approaches [60]. Comprehensive indicators can more finely reveal how different components within an ecosystem interact and influence final outcomes[61-62].

In service ecosystems, the concepts of entropy and efficiency have become key metrics for assessing system complexity and dynamism. Entropy, derived from information theory, quantifies the disorder or unpredictability within a system[63]. Higher entropy values indicate increased uncertainty in system behavior, which is particularly significant in the context of swarm intelligence, as diverse individual behaviors and complex group dynamics can lead to heightened system disorder [64-65]. Research shows that increased system entropy is often associated with the diversity of interactions among system components, which frequently give rise to emergent behaviors that traditional frameworks fail to capture [66].

In the nonlinear context, the definition of efficiency has been expanded to include not only resource utilization but also adaptability and long-term stability. These attributes are critical for effective service delivery in rapidly changing environments [67]. For instance, by measuring a system's adaptive capacity and resilience, it becomes possible to better evaluate its performance in responding to evolving challenges.

In practical applications, the integration of methods such as Data Envelopment Analysis (DEA) and Key Performance Indicator (KPI) frameworks can provide comprehensive performance evaluations for service ecosystems. These methods address the complexity introduced by nonlinear interactions by analyzing multiple variables [68] . However, these approaches also face significant challenges, particularly in terms of data integration and the complexity of coordinating across stakeholders[69].

Despite the immense potential of nonlinear metrics to enhance the understanding of service ecosystems, their application remains fraught with challenges. Complex data collection, difficulties in model construction, and the intricacies of causal interpretation all require further research [70-71] . Future studies should delve deeper into the roles of entropy and efficiency across different contexts to advance the realization of sustainable and resilient service ecosystems, ensuring their ability to effectively adapt to changing demands and pressures.
\section{Analysis of Service Ecosystems}
Analyzing service ecosystems has become increasingly challenging, primarily due to the self-organizing nature of collective intelligence. This characteristic leads to complex higher-order dynamic behaviors that are difficult to capture and analyze using traditional quantitative methods. Furthermore, the adaptive changes of individual agents and the secondary emergence of collective behaviors exacerbate system complexity, posing significant challenges for conventional analytical approaches in understanding and predicting system dynamics[72].

To gain deeper insights into service ecosystems, it is essential to adopt a collective intelligence perspective and explore multiple dimensions that reveal the system's complexity and underlying principles. These dimensions not only provide unique viewpoints but also enhance our understanding of the system's dynamic behaviors and characteristics.
\begin{figure}[!t]
    \centering
    \includegraphics[width=0.6\linewidth]{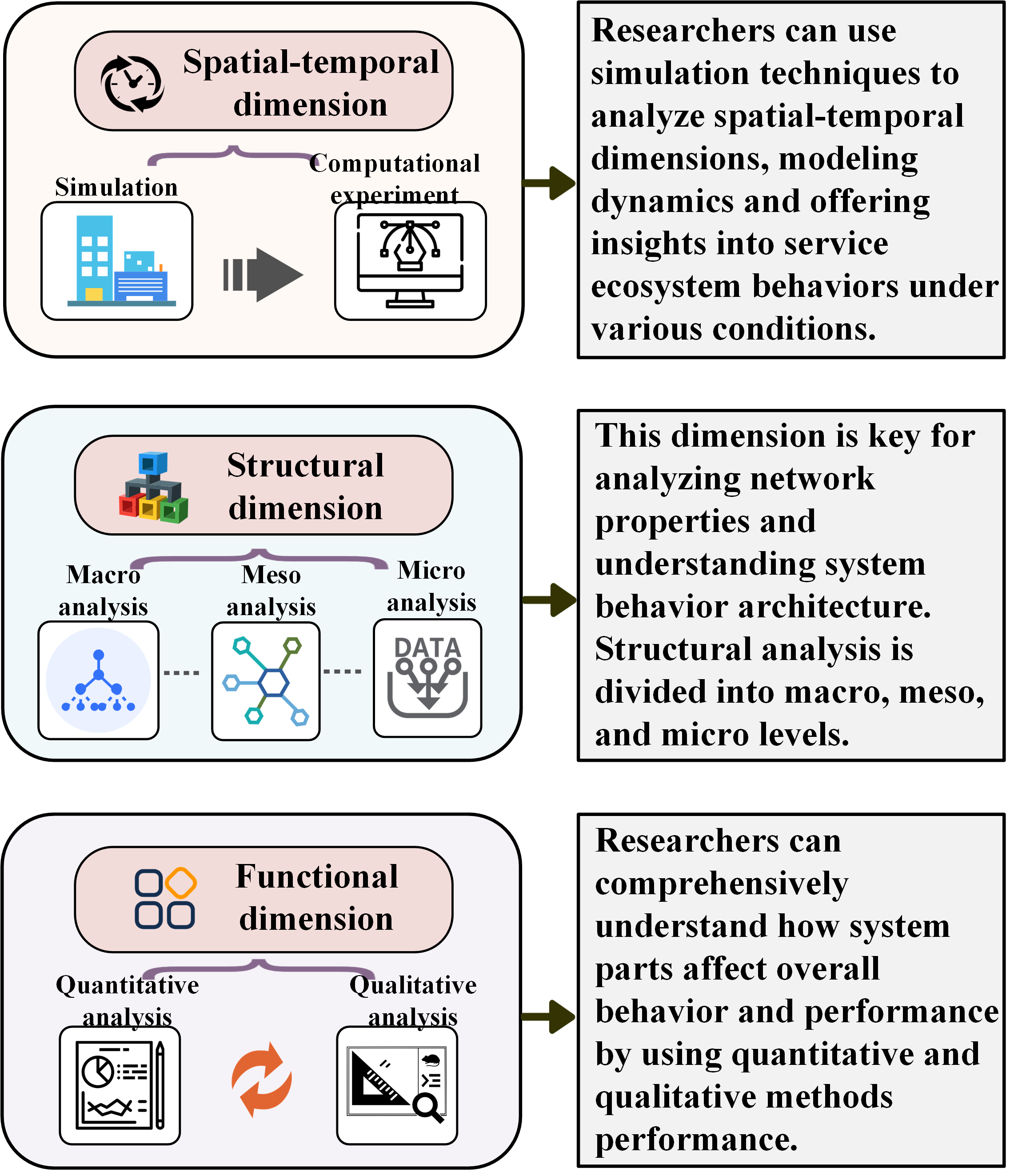}
    \caption{Spatial, temporal, structural and functional dimensions of service ecosystem analysis.}
    \label{fig_1}
\end{figure}

\subsection{Multidimensional Analysis of Swarm Intelligence-Driven Service Ecosystems}
Before the advent of collective intelligence, service systems primarily relied on traditional analytical methods, such as queuing theory and statistical models[73], which centered on centralized control and predictive modeling. These methods often assumed fixed parameters, making them ill-suited to handle uncertainties in complex, dynamic environments[74]. For instance, queuing theory optimizes service operations by studying metrics like waiting times and service rates [75-76]. However, traditional approaches fail to effectively capture the core characteristics of collective intelligence systems—self-organization, decentralization, and emergent behaviors—resulting in limitations when interpreting the true dynamics of such systems.

The complexity and dynamism of service ecosystems have surpassed the capabilities of traditional methods. Therefore, it is necessary to adopt a collective intelligence perspective and analyze these systems through three dimensions: temporal-spatial, structural, and functional.
\begin{itemize}
    \item \textbf{Spatial-temporal dimension:} Analyzing the spatial and temporal interactions among agents within service ecosystems reveals how these interactions form dynamic and complex networks, and explores the impact of their evolutionary processes on the system's overall behavior. For instance, the spatiotemporal distribution of service requests influences resource allocation and system efficiency.
    \item \textbf{Structural dimension:} Investigating the organizational structure and hierarchical relationships of service ecosystems reveals the behavioral mechanisms and evolutionary patterns at macro, meso, and micro levels. For instance, understanding how individual collaboration mechanisms influence overall system performance is critical to uncovering the dynamics of service ecosystems.
    \item \textbf{Functional dimension:} Analyzing the roles of system components and their interaction patterns reveals how individual behaviors, through local interactions, give rise to complex global behaviors. For instance, the decision-making rules of service agents significantly influence the effectiveness of service delivery.
\end{itemize}

This multi-dimensional analytical approach not only uncovers the dynamic behaviors and characteristics of the system but also provides a critical theoretical foundation for system optimization.
\subsection{Spatiotemporal Analysis in Service Ecosystems}
In service ecosystems, agents interact across space and time, forming a dynamic and complex network. The behavior and evolution of this network can be viewed as a manifestation of collective intelligence, where individuals, through local interactions and simple rules, give rise to complex global behaviors. Simulation techniques enable researchers to model the system's dynamic behaviors, providing valuable insights into its performance under varying conditions. Computational experiments, as a robust research method, leverage simulation and modeling to deeply analyze system behavior and performance, offering essential tools for understanding the emergent mechanisms of collective intelligence in service ecosystems[77-78].

Analyzing the spatiotemporal dimension in collective intelligence-driven service ecosystems presents significant challenges due to their dynamic and complex nature. Raudsepp-Hearne et al.highlight that intricate interactions within service ecosystems make their value difficult to quantify accurately [79], while Liu et al.demonstrate that spatial distribution, influenced by human intervention, further complicates system analysis [80]. By employing multi-agent simulation (ABM) and stakeholder engagement methods, combined with theoretical frameworks of collective intelligence and advanced spatiotemporal modeling techniques [81], a deeper understanding of individual and collective behaviors in complex systems can be achieved, paving the way for practical and inclusive solutions for sustainable governance[82].

Future research should focus on developing more sophisticated spatiotemporal analytical frameworks to elucidate the nonlinear relationships between service ecosystems and the socioeconomic factors driving their changes [83-84].
\subsection{Structural Dimension Analysis}
From the perspective of collective intelligence, analyzing the structural dimension of service ecosystems is crucial for understanding their complexity and dynamic behaviors. At its core, collective intelligence involves individuals generating complex global behaviors through local interactions and simple rules, with the structural characteristics of service ecosystems serving as the foundation for such emergent behaviors. Structural analysis can be conducted at macro, meso, and micro levels, each revealing the behavioral mechanisms and evolutionary patterns of the system at different scales.

\textbf{Macro analysis (top-down): } Focusing on the overall structure and behavior of service ecosystems, this approach aims to uncover the driving forces, evolutionary mechanisms, and current operational states through systems thinking and data analysis. Watson R et al. emphasize the importance of comprehensive evaluation frameworks in capturing interactions among ecosystem services, demonstrating that such innovative methods enhance the management and understanding of ecological functions [85]. Zhu W et al. propose a conceptual framework capable of effectively integrating various aspects of ecosystem services, with a particular focus on the supply and demand of services, thereby revealing broader behavioral patterns [86]. Macro-level analysis provides critical insights into the global behavior of service ecosystems, elucidating their overall structure and evolutionary mechanisms.

\textbf{Meso analysis (linking macro and micro): }Serving as a bridge between macro and micro perspectives, this approach focuses on the interactions among nodes within sub-networks of service ecosystems and their structural properties. It not only emphasizes the predictive power of interactions but also facilitates rapid problem identification and resolution. Research at the meso level reveals how individual interactions aggregate to form the collective behavioral characteristics of collective intelligence. Zhu W et al. highlight that integrating collective optimization techniques into power communication systems demonstrates how local decision-making enhances overall system efficiency. By identifying and analyzing these interactions, organizations can leverage the inherent advantages of collective behaviors to optimize responses to dynamic operational conditions and strengthen the resilience of service delivery processes. The meso perspective provides a deeper understanding of how individual behaviors influence the overall system.

\textbf{Micro analysis (bottom-up): }At the individual level, the evolution of service ecosystems is examined by highlighting the behaviors and interactions of multiple agents. Micro-level analysis offers a detailed view of how individual adaptive changes influence the broader ecosystem. Although individual agents exhibit heterogeneous behaviors, their collective interactions give rise to emergent patterns that impact the larger service ecosystem. The complexity of these dynamics necessitates robust service discovery and resource allocation mechanisms capable of adapting to the evolving behaviors and interactions of individual agents [87].

Despite the advantages of structural dimension analysis, this approach is not without limitations. The risk of oversimplification becomes particularly pronounced when complex interactions are reduced to discrete hierarchical levels, potentially obscuring the relevance and fluidity of inter-level relationships [88] . Furthermore, the inherent dynamism and nonlinearity of collective intelligence systems pose significant challenges to traditional hierarchical analysis methods, necessitating the exploration of more integrated analytical frameworks [89] . Future research should focus on developing comprehensive frameworks that account for the dynamic interactions between these levels while capturing the multi-dimensional characteristics of service ecosystems.
\subsection{Functional Analysis in Service Ecosystems}
Functional dimension analysis examines the roles and interactions of various components within a system, revealing how individual behaviors in collective intelligence give rise to complex global behaviors through local interactions. This analysis employs both quantitative and qualitative methods to provide a holistic view of how different elements influence overall system behavior and performance. Mixed methods are particularly advantageous for informing decision-making and developing guidelines for complex interventions, as they acknowledge the importance of context and the intricate interactions among system components [90-91].

Through systems thinking and system dynamics modeling, combined with precise data computation and complex operations, the relationships and feedback loops within ecosystems can be uncovered [92-93]. This approach quantifies system behavior and aids in understanding its underlying mechanisms. However, when confronted with the emergent behaviors of collective intelligence, system dynamics modeling may struggle to fully capture the system's nonlinear dynamics and individual adaptive changes. For instance, service providers adjusting strategies based on market demand can trigger global changes in resource allocation through local interactions, a complex dynamic process that traditional quantitative methods may not fully quantify.

Qualitative analysis focuses on constructing models that describe system structures, relationships, and behavioral patterns without relying on precise numerical data [94]. This method can reveal nuances and multidimensional characteristics that quantitative analysis might overlook, such as the diversity of individual behaviors and the emergence of collective behaviors. Qualitative analysis offers unique advantages in capturing the self-organizing nature of collective intelligence, providing deeper insights into the dynamic evolution of systems. Mixed methods combine the strengths of both quantitative and qualitative approaches to offer a more comprehensive understanding of ecosystem services [95] . For example, by integrating quantitative analysis of feedback loops with qualitative analysis of subtle behavioral variations, emergent behaviors in collective intelligence can be better explained.

However, the self-organizing nature of collective intelligence makes integrating findings from these two methods more challenging[96-97]. Specifically, reconciling precise data from quantitative analysis with behavioral patterns from qualitative analysis poses a pressing issue in ensuring that research outcomes are actionable for policymakers and practitioners [98].
\section{Optimization of the Service Ecosystem}
In modern service systems, individual agents coalesce into complex service networks through principles of collective intelligence, with decentralized structures enhancing flexibility and adaptability, thereby driving innovation across multiple domains. However, autonomous decision-making by individual agents based on local information often leads to misalignment between individual behaviors and system-wide objectives, particularly in terms of consistency and coordination, resulting in resource inefficiencies and system instability. The emergent behaviors of collective intelligence further complicate optimization, rendering traditional centralized governance models inadequate and necessitating the development of governance frameworks that reconcile individual autonomy with collective goals.

Optimization of service systems can be achieved through direct methods, such as dynamic pricing and individual incentives to enhance operational efficiency and revenue generation, and indirect methods, leveraging real-time feedback and machine learning to bolster adaptive capabilities. While these approaches differ in focus, both aim to balance individual and collective interests, improving system competitiveness and operational efficiency in dynamic environments.
\begin{figure*}[htbp]
    \centering
    \includegraphics[width=0.9\textwidth]{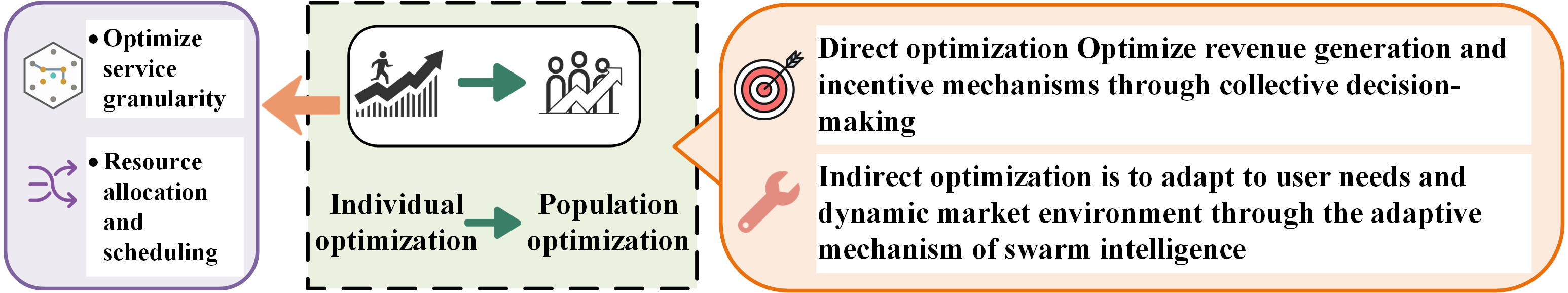} 
    \caption{Optimization diagram of collective intelligence in service ecosystem.}
    \label{fig:optimization_diagram}
\end{figure*}

\subsection{The Dilemma of Collaborative Optimization in Swarm Intelligence}
In modern service systems, individual agents form complex service networks through principles of collective intelligence, with decentralized structures significantly enhancing flexibility and adaptability, driving innovative solutions in fields such as robotics, artificial intelligence, and distributed computing [99]. However, the inherent collaborative behaviors of collective intelligence pose significant challenges to global optimization. At its core, individual agents make autonomous decisions based on local information and simple behavioral rules, akin to natural swarm phenomena observed in insects like ants and bees [100]. While this decentralized approach enhances adaptability, it often leads to misalignment between individual behaviors and system-wide objectives, particularly in terms of consistency and coordination [101]. For instance, agents prioritizing local efficiency may inadvertently overlook collective goals, resulting in conflicts between individual and system interests. Empirical studies in dynamic environments, such as shared economies or distributed energy systems, reveal that excessive focus on local optimization can lead to resource waste and system instability. When agents neglect the broader impacts of their actions on resource allocation fairness and system sustainability, achieving global optimization becomes a significant challenge [102].

Moreover, the emergent behaviors characteristic of collective intelligence add another layer of complexity to optimization efforts. Nonlinear and dynamic interactions among agents often produce outcomes that are difficult to predict using simple analytical models, rendering traditional centralized governance models ineffective. For example, hierarchical management practices in decentralized networks, such as shared resource platforms, often fail to address inefficiencies, exacerbating congestion and underutilization [103]. Consequently, there is an urgent need to develop governance frameworks and optimization mechanisms that reconcile collective intelligence principles with individual agent autonomy. Such frameworks must balance dynamic and often competing interests within the system, ensuring operational efficiency aligns with collective goals. This remains a critical area of research with significant implications for the design and management of modern service systems.

Governance and optimization of service systems can be achieved through two primary approaches: direct optimization via incentives and indirect optimization through adaptive mechanisms. Direct optimization involves explicit reward structures to motivate individual agents, enhancing performance and productivity. Indirect optimization relies on real-time feedback and dynamic adjustments, leveraging the system's self-regulating capabilities to adapt to environmental changes and user needs. Both approaches, though distinct, are essential for optimizing service systems, enabling organizations to maintain competitiveness and operational efficiency in complex and dynamic environments.
\subsection{Direct Optimization Based on Incentive Mechanisms}
From the perspective of collective intelligence, direct optimization through revenue generation and incentives reveals the significant role of collective decision-making among individual agents in enhancing operational efficiency, formulating adaptive pricing strategies, and improving service outcomes. The core of collective intelligence lies in its decentralized decision-making mechanism, where individual agents make autonomous decisions based on local information, a feature prominently reflected in dynamic pricing strategies. For instance, dynamic pricing mechanisms employed by companies like Uber and Lyft demonstrate how collective adjustments based on consumer behavior can optimize service delivery while boosting overall revenue [104]. The effectiveness of this mechanism stems from its ability to allow individual agents to rapidly respond to local environmental changes, benefiting the entire system by attracting more users and maximizing profits. Furthermore, studies show that aligning individual incentives with group performance significantly enhances the collective efficiency of service systems. For example, Abhishek et al. highlight that linking incentives to service quality and performance metrics not only motivates individual performance but also optimizes resource allocation and operational efficiency [105].

On the other hand, collective intelligence achieves resource optimization through collective adaptation, demonstrating notable advantages in dynamic environments. Ocran et al. illustrate that optimizing domestic tourism demand shifts using dynamic pricing models significantly improves economic outcomes, indicating that collective pricing strategies based on adaptive mechanisms yield better economic results [106]. Additionally, the application of collective intelligence methods, such as Particle Swarm Optimization (PSO), in dynamic pricing further enhances decision-making efficiency. Sakurama notes that dynamic pricing optimization based on real-time supply-demand interactions effectively adjusts station locations and pricing strategies, maximizing operational efficiency and user satisfaction [107]. These methods enable agents to collectively optimize prices based on local data through self-organizing behaviors, showcasing the broad potential of collective intelligence in service systems.

By implementing dynamic pricing schemes, aligning individual incentives with group objectives, and adopting adaptive resource optimization strategies, service systems can achieve superior operational outcomes and profitability, underscoring the critical role of collective intelligence in service system governance. Future research on direct optimization through incentives should focus on real-time feedback, personalized incentive design, and the evaluation of long-term incentive effects to enhance precision and sustainability.
\subsection{Indirect Optimization Based on Adaptive Mechanisms}
Collective intelligence emphasizes decentralized, self-organizing system behaviors, enabling the resolution of complex problems through the collaboration of multiple simple agents. This approach not only enhances the operational efficiency of service systems but also improves their adaptability to user needs and dynamic market conditions. For instance, Wu et al. demonstrated that intelligent customer service platforms optimize user interactions through real-time feedback mechanisms, allowing the system to self-adjust based on user behavior and preferences, thereby delivering dynamic services [108]. Such adaptive capabilities enable service systems to remain relevant in complex and rapidly changing environments, particularly when consumer demands evolve quickly. Nilashi M further highlights that optimizing feedback systems can strike a balance between rapid decision-making and controlled moderation, thereby improving service quality and user satisfaction [109] .

Moreover, metacognitive capabilities within collective intelligence play a pivotal role in service system governance. Metacognition, defined as the awareness and regulation of one's cognitive processes, manifests at the group level as collective reflection and continuous improvement. For example, through shared experiences and mutual learning, individual agents in a group can draw lessons from successful decisions, enhancing overall performance [110]. This collective learning mechanism not only boosts individual agent efficiency but also strengthens the collective intelligence of the service system. Additionally, the integration of machine learning technologies further amplifies the system's adaptive capabilities. By analyzing historical data and dynamically adjusting operational strategies, intelligent systems can better meet user needs. For instance, in logistics and customer service, machine learning algorithms predict future demands and proactively adjust service strategies, optimizing resource allocation [111] .

Adaptive mechanisms based on collective intelligence not only improve the responsiveness and service quality of systems but also achieve higher operational efficiency through collective intelligence and resource optimization, providing critical support for modern service system optimization. Research on indirect optimization via adaptive mechanisms should focus on the development of intelligent systems, the application of collective intelligence, and empirical validation to enhance the dynamic adaptability and robustness of service systems.
\section{Service Ecosystem: A Case Study of Ride-Sharing Platforms}
Shared mobility platforms integrate diverse resources, technologies, and participants (e.g., drivers, passengers, and third-party service providers) to form a complex and interconnected service ecosystem. Traditionally, linear metrics such as response time, passenger wait time, and driver acceptance rate have been central to evaluating system efficiency. However, these metrics often fall short in addressing the growing complexity of ride-hailing systems, which are significantly influenced by dynamic factors like traffic conditions and weather, impacting passenger wait times and supply-demand matching [112-113]. This limitation has prompted a shift toward nonlinear metrics, such as entropy and efficiency, which better capture the underlying complexities of ride-hailing environments. For instance, entropy measures the uncertainty in passenger-driver matching, while efficiency evaluates the adaptability of resource allocation [114-115] .

The three-dimensional complexity of ride-hailing systems—temporal-spatial dynamics, structural considerations, and functional interactions—further complicates performance measurement. Ride demand fluctuates with time and geographic regions, peaking during morning and evening commutes [116] . By leveraging spatiotemporal analysis, platforms can predict these fluctuations and proactively adjust pricing and dispatching strategies to balance supply and demand [117]. Structural analysis helps identify organizational hierarchies, from individual driver behavior to broader system optimization [118]. Functional analysis reveals how local interactions between drivers and passengers collectively shape overall system behavior; for example, individual driver decisions on order acceptance can significantly influence supply-demand patterns [119].

Additionally, conflicts of interest between individual drivers and collective passenger needs add another layer of complexity [120]. Drivers are often incentivized to prioritize high-revenue orders, potentially neglecting low-demand areas with unmet passenger needs. This misalignment necessitates both direct and indirect optimization strategies. Directly, platforms implement dynamic pricing mechanisms, such as Uber's surge pricing, to encourage driver participation during peak demand [121]. Indirectly, real-time feedback systems and machine learning algorithms optimize routes and dispatching; for instance, DiDi employs data-driven approaches to minimize passenger wait times and driver idle periods [122] .

Future research could explore the application of these frameworks to swarm intelligence in various service ecosystems, potentially enhancing efficiency and responsiveness in fields like logistics and public transportation [123].
\section{Conclusion}
This paper explores the complexity of service ecosystems and the challenges associated with their governance, particularly through the lens of swarm intelligence. By dividing service ecosystems into three evolutionary stages—biologically inspired intelligence (Swarm Intelligence), socially inspired intelligence (Collective Intelligence), and brain-inspired intelligence (Crowd Intelligence)—it provides a comprehensive framework that captures the developmental trajectory and inherent patterns of collective intelligence in these systems. This theoretical framework not only elucidates the mechanisms of collective intelligence formation but also offers new perspectives for understanding the complexity of service ecosystems.

The primary contribution of this paper is the proposal of a cyclical analysis framework for collective intelligence, encompassing three dimensions: "Measurement-Analysis-Optimization." This framework emphasizes the cyclical evolution mechanisms within service ecosystems, including measurement based on linear and nonlinear perspectives, analysis considering temporal-spatial, structural, and functional dimensions, and optimization grounded in individual and collective aspects. By integrating these elements, our framework provides new methodological support for studying the evolutionary mechanisms of service ecosystems.

The paper also conducts an in-depth analysis of the service ecosystem of ride-sharing platforms, revealing the limitations of traditional linear metrics in addressing system complexity and proposing nonlinear metrics (such as entropy and efficiency) to more comprehensively reflect the dynamic characteristics of the ride-hailing environment. Through analysis across temporal-spatial, structural, and functional dimensions, the paper investigates the impact of demand fluctuations, system hierarchies, and local interactions on overall behavior. It further proposes direct and indirect optimization strategies, such as dynamic pricing and real-time feedback systems, to address conflicts of interest between drivers and passengers. These findings not only provide theoretical support for optimizing ride-sharing platforms but also offer new perspectives for the application of swarm intelligence in other service ecosystems, potentially enhancing efficiency and responsiveness in fields such as logistics and public transportation.

The paper underscores the importance of understanding the construction process of collective intelligence systems, which can provide a solid theoretical foundation for subsequent ecological research in areas such as supply-demand matching, value realization, collaborative services, and governance. Additionally, it highlights the need for further research on the regulatory governance of service ecosystems, aiming to contribute to the formation of a stable, orderly, and dynamic knowledge system for service governance.

\end{document}